\documentclass[twoside]{book}
\usepackage{amsmath}
\usepackage{graphicx}
\usepackage{latexsym}
\usepackage{amsfonts}
\usepackage{amssymb}
\usepackage{bm}
\usepackage{amsbsy}
\usepackage{longtable}
\usepackage{wrapfig}
\numberwithin{equation}{chapter}
\usepackage{makeidx}
\makeindex

\begin{document}

\thispagestyle{empty}

\begin{centering}

\Large{Polymer Physics:}\\

\vspace{2ex}

\Large{Phenomenology  of}\\

\vspace{2ex}

\Large{Polymeric Fluid Simulations}

\vspace{2ex}

\Large{George D. J. Phillies}\\

\large{Professor of Physics, Emeritus}\\

\large{Worcester Polytechnic Institute}

\end{centering}

\renewcommand{\thepage}{\roman{page}}

\setcounter{chapter}{0}
\setcounter{page}{1}
\renewcommand{\thechapter}{\arabic{chapter}}
\renewcommand{\thepage}{\arabic{page}}
\chapter{Collective Coordinates and Collective Models \label{COORDcollective} }

\section{Introduction \label{COORDintroduction}}

This Chapter considers collective coordinates and collective models for isolated polymer chains. Collective coordinates can be effective tools for isolating aspects of polymer motion that are less obvious if only the Cartesian coordinates of the individual atoms or monomers are viewed. One of the sets of collective coordinates that we consider, the Rouse coordinates, follows naturally from the Rouse model of polymer dynamics.  We therefore consider the Rouse model
as well as its more important predecessor, the Kirkwood-Riseman model

Collective coordinates are generated as weighted sums, or more complicated numerical processes, over the coordinates of the atomic or monomer coordinates. In this chapter, we consider three sets of collective coordinates, namely Fourier coordinates, Rouse coordinates\cite{COORDrouse1953a,COORDzimm1956a}, and Haar wavelet coordinates\cite{COORDhaar1910a,COORDdaubechies1992a}.
Fourier coordinates arise naturally in any discussion of scattering, whether by light, x-rays, or neutrons, by polymer solutions; they also rise naturally in discussions of pulsed-gradient spin-echo nuclear magnetic resonance, though the nominal wave vector $\mathbf{q}$ in PGSE NMR has an entirely different physical origin and meaning than does the scattering vector $\mathbf{q}$ in a scattering experiment.  Rouse coordinates appear naturally as solutions of the Rouse and Zimm models.  Each Rouse or Fourier coordinate represents a weighted sum over the positions of all the atoms or monomers in the polymer chain.  Sometimes one might wish to consider instead the dynamics of a limited segment of a larger chain. For this purpose Haar wavelets represent a natural approach to treating segmental dynamics, and correspond to directly to studies of block copolymers using dielectric relaxation spectroscopy.

Here we make a careful distinction between \emph{coordinates}, \emph{components} or \emph{amplitudes}, and \emph{modes}.  The \emph{coordinates} are a list of variables that jointly specify, when their numerical values are supplied, the positions of all the particles in the system.  For example, for particle $1$ the Cartesian position coordinates might reasonably be denoted by the algebraic symbols $(x_{1}, y_{1}, z_{1})$.  The Cartesian\emph{ components} for particle 1 at a particular time $t$ are the numerical values that the position coordinates happen to have at time $t$.  For example, $(x_{1}(t), y_{1}(t), z_{1}(t)) = (0,0,0)$ corresponds to a particle that is located at the origin at time $t$.  The term \emph{amplitudes} rather than\emph{ components} is more obviously appropriate when one considers Fourier transforms.  The particle position at time t may be viewed as a position vector $\mathbf{R}_{1}(t) = (x_{1}(t), y_{1}(t), z_{1}(t))$; for the same particle the components of that position vector are $(0,0,0)$.

Finally, a normal of the system is a set of coordinates whose evolution in time does not change the ratio of the amplitudes of the coordinates.  For a relaxational mode, if $\mathbf{A}(t)$ gives the amplitude of a mode at time $t$, then
\begin{equation}
    \mathbf{A}(t) = \mathbf{A}(0) f(t),
    \label{eq:COORDmode}
\end{equation}
where $f(t)$ is a function of time determined by the mode under consideration. Modes are said to be \emph{independent} if the thermal fluctuations in their values are uncorrelated. It is not necessarily the case that polymer motions can be decomposed into a complete set of independent modes that jointly describe all possible motions of a polymer.

As a familiar example of eq.\ \ref{eq:COORDmode}, the vibrational modes of a taut undamped string having fixed ends are
\begin{equation}
   y(x,t)  =  [A cos(kx+\beta)] [sin(\omega t+ \alpha)]
   \label{eq:COORDstringmodes}
\end{equation}
In this equation, $y$ is the lateral displacement of the string, $x$ is the field variable denoting the position along the string, $\beta$ and $\alpha$ are phase factors, and $A$ is the \emph{amplitude} of the oscillation. Also, $k$ and $\omega$ are the wave vector and angular frequency of the vibrational mode.  The first quantity in square brackets represents the spatial dependence of the mode.  The second quantity in square brackets represents the oscillating temporal dependence $f(t)$ of the mode.

As descriptions of the positions of the atoms in a polymer, Fourier and Rouse coordinates have in common that they depend on the relative positions of more or less all the atoms.  In some cases, one refers to groups of atoms, united atoms or monomers or beads, rather than single atoms, but what is said for atomic positions remains true for the positions of united atoms or monomers or beads.  We say 'more or less all the atoms' because the transformations from Cartesian to Fourier or Rouse coordinates sometimes have isolated zeros, so that occasionally a particular atom's Cartesian coordinates do not contribute to a particular Fourier or Rouse component.  Fourier and Rouse coordinates each have a scale parameter (for Fourier components, this is the magnitude $k$ of the wavevector of the coordinate) that reflects the smaller or larger distance over which the coordinate reflects relative positions.  However, even for a short-distance scale parameter, the component associated with a given Fourier or Rouse coordinate is determined by the relative positions of all the atoms under consideration.  Fourier and Rouse coordinates may reflect either short- or long-distance correlations, but they are not localized coordinates; they always reflect the positions of all the atoms in a chain.  In contrast, the Haar wavelet components described below are localized.  They reflect the relative positions of atoms in short or long sections of a chain, but do not (with one exception, the chain center-of-mass coordinate) depend on the positions of every single atom in a polymer chain.

Some sets of coordinates are said to be \emph{complete} or \emph{orthogonal}.  A set of coordinates  is  \emph{complete} if every point in the system can be represented by a set of values for the coordinates.  For example, the Cartesian $x$, $y$ and $z$ coordinates are complete, in that every point in space has a set of numerical values for its $x$, $y$, and $z$ coordinates, and every set of $x$, $y$, and $z$ components corresponds to a point in space.  Incomplete sets of coordinates are often of practical theoretical value.  For example, in the original Rouse model, polymer beads represent averages over substantial numbers of neighboring monomers, while Rouse-model polymer springs represent polymer strands that are long enough that their end-to-end distances were adequately described by Gaussian statistics.  In the original Rouse model the coordinates of atoms within a single bead or spring were not represented by Rouse coordinates, meaning that the Rouse coordinates are an incomplete set of coordinates.  Some sets of coordinates are \emph{overcomplete}, so that the same point in space is represented by several different sets of coordinates.  Circular polar coordinates, in which angular coordinate components separated by $2n \pi$ refer to the same point, are an example of overcomplete coordinates.  This neglect of atomic coordinates within the Rouse model, in favor of course grained coordinates, has consequences discussed elsewhere if for some reason one wants to use the model to make quantitative predictions.

What is meant by \emph{orthogonal}?  For a Cartesian vector $\mathbf{R}$,
\begin{equation}
   \mathbf{R} = a \mathbf{\hat{i}} + b \mathbf{\hat{j}}  +  c \mathbf{\hat{j}},
   \label{eq:COORDcartesian}
\end{equation}
in which $a$, $b$, and $c$ are the values of the $x$, $y$, and $z$ coordinates and $\mathbf{\hat{i}}$, $\mathbf{\hat{j}}$, and $\mathbf{\hat{k}}$ are the standard Cartesian unit vectors, orthogonality is defined by the vector dot product (the \emph{scalar} product)  and the relations
\begin{eqnarray}
   \mathbf{\hat{i}} \cdot \mathbf{\hat{i}} &=& 1, \\
   \mathbf{\hat{j}} \cdot \mathbf{\hat{j}} &=& 1, \\
   \mathbf{\hat{k}} \cdot \mathbf{\hat{k}} &=& 1, \\
   \mathbf{\hat{i}} \cdot \mathbf{\hat{j}} &=& 0, \\
   \mathbf{\hat{j}} \cdot \mathbf{\hat{k}} &=& 0, \\
   \mathbf{\hat{k}} \cdot \mathbf{\hat{i}} &=& 0. \\
   \label{eq:COORDdotproduct}
 \end{eqnarray}

In order to treat orthogonality of other sorts of coordinates, one first needs to find a definition of a scalar product.  For physical quantities measured at the same time in an equilibrium system, a reasonable definition for the scalar product is the equilibrium canonical ensemble average.    For two quantities $B$ and $C$, that average is
\begin{equation}
      \langle B  C \rangle   =  \int d \mathbf{\Gamma} B(\mathbf{\Gamma}) C(\mathbf{\Gamma})  \exp(- \beta (H(\mathbf{\Gamma})-A).
     \label{eq:COORDcorrelation}
\end{equation}
In this equation, $B$ and $C$ are the quantities being averaged.  $\mathbf{\Gamma}$ represents all the position and momentum coordinates, so the indicated integral is actually $6N$-dimensional, $N$ being the number of particles in the system. $\beta = 1/(k_{B} T)$, $k_{B}$ being Boltzmann's constant and $T$ being the absolute temperature.  Finally, $H(\mathbf{\Gamma})$ is the energy of the system for a given set of values of the positions and momenta, while $A$ is the Helmholtz free energy.  The physical quantities $B$ and $C$ are orthogonal (or \emph{uncorrelated}) if $\langle B C \rangle =0$.  Some authors, to encourage the notation to resemble more closely the Dirac bra-ket notation of quantum mechanics, will insert a complex conjugate into the ensemble average, by writing the scalar product of $B$ and $C$ as $\langle B^{*} C \rangle$.

One may readily average a product of more than two quantities, as treated when we discuss spatial Fourier components of the density. In a non-equilibrium system, such as a liquid subject to shear flow, a reasonable definition of the scalar product becomes more complicated.

For various reasons, it is also of interest to consider time correlation functions, in which one calculates an average over quantities measured at two different times $t$ and $t+\tau$, e.g.
\begin{equation}
    C(\tau) =  \langle A(t) B(t+\tau) \rangle.
    \label{eq:COORDtcf}
\end{equation}
In an equilibrium system, $C$ is independent of the clock time $t$; it depends only on the time separation $\tau$. In non-equilibrium systems having time evolution, matters become more complicated; C may then depend both on $t$ and on $\tau$. In this case, avail may be had, e.g., to bispectral analysis \cite{COORDphillies1980a}.  In calculations of correlation functions, the brackets $\langle \cdots \rangle$ may denote either an ensemble average or a time average.  In taking the ensemble average, the fact that the system is deterministic may be used to write $B(t+\tau)$ in terms of the positions and momenta of all the beads at time $t$ (or vice versa for $A(t)$), in which case $C(\tau)$ reduces to a simple canonical ensemble average over a highly complex integrand. In simulations, advantage is taken of the general feature that, in the systems under consideration here,  ensemble and time averages lead --- up to fluctuations --- to the same answer, so in evaluating $C(\tau)$ via a time average what is done is to calculate $A(t)$ and $B(t)$, each at a long series of times, and then cross multiply $A$ at each time $t$ with $B$ calculated at each time $t+\tau$.

\section{Fourier Components \label{COORDfourier}}

The simplest example of a Fourier transform may be written
\begin{equation}\label{eq:fourierexample}
    F(k) = \int_{-\infty}^{\infty}  dx f(x) \exp(\imath k x).
\end{equation}
Here $f(x)$, which is a function of $x$, is the function being transformed. $F(k)$, which is a function of $k$, is the Fourier transform of $f(x)$. The variables $x$ and $k$ are the \emph{dual variables} in the transform, the transform replacing $x$ with $k$.   The inverse transform, an integral of $F(k)$ over all values of $k$, replaces $k$ with $x$ and transforms $F(k)$ back into $f(x)$.  Because the above equation shows the continuous transform, between the transform and its inverse there is a normalizing factor $2 \pi$ which different authors distribute differently between the transform and its inverse.

Here the transform coordinate is $k$, the amplitudes for the coordinate $k$ are the $F(k)$, and a hypothetical mode --- if one existed --- would correspond to the time dependence of $F(k,t)$.

The Fourier components of the concentration of polymer beads are of broad general interest, for example because Fourier components represent modes of the simplest diffusion equation. For a set of $N$ beads having time-dependent locations $\mathbf{r}_{i}(t)$, the spatial Fourier components of the concentration are
\begin{equation}
   a_{\mathbf{q}}(t)  = \sum_{j=1}^{N} a_{i} \exp(-\imath \mathbf{q} \cdot \mathbf{r}_{i}(t))
   \label{eq:COORDfouriera}
\end{equation}
Here the dual variables of the transform are the position $\mathbf{r}$ and the spatial Fourier transform vector $\mathbf{q}$.  Equation \ref{eq:COORDfouriera} is a discrete transform. The function $a_{i}$ being transformed only exists at a list of values $\mathbf{r}_{i}(t)$, so $\mathbf{r}$ has already been replaced with its values $\mathbf{r}_{i}(t)$.  In light, x-ray, and neutron scattering calculations, $\mathbf{q}$ is the scattering vector, as discussed in standard texts.  The same form, with an entirely different meaning for $\mathbf{q}$, arises in treatments of pulsed-gradient spin-echo NMR measurements.   The $a_{i}$ reflect the possibility that different beads have different scattering powers and hence different statistical weights.

Fourier components are said to be orthogonal, but two different meanings of \emph{orthogonal} should be kept in mind.  First
\begin{equation}
    \int^{\infty}_{-\infty} dx \ \exp(\imath q x)  \exp(-\imath k x) = 0
    \label{eq:COORDsimpleorthogonal}
\end{equation}
if $k \neq q$.  That's the math result for a plausible set product of two Fourier coordinates.  Second, for beads in a fluid, $\langle a_{\mathbf{q}}(t) \rangle$ vanishes by translational invariance, namely if one relocates the origin by $\mathbf{O}$, a spatial Fourier component becomes
\begin{equation}
   a_{\mathbf{q}}(t)  = \sum_{j=1}^{N} a_{i} \exp(-\imath \mathbf{q} \cdot (\mathbf{r}_{i}(t)-\mathbf{O})),
   \label{eq:COORDfourierb}
\end{equation}
so as a result of moving the origin $\langle  a_{\mathbf{q}}(t) \rangle$ becomes $\langle  a_{\mathbf{q}}(t) \rangle \exp(\imath \mathbf{q} \cdot \mathbf{O})$.  However, there is no physical meaning to the location of the origin, because the system has translational invariance, so changing $\mathbf{O}$ can have no effect on $\langle a_{\mathbf{q}}(t) \rangle$, which is only possible if $\langle a_{\mathbf{q}}(t) \rangle = 0$.

A typical scattering experiment determines the intermediate structure factor $g^{(1)}(\mathbf{q}, t)$, namely
\begin{equation}
   g^{(1)}(\mathbf{q}, \tau) = \langle a_{\mathbf{-q}}(t) a_{\mathbf{q}}(t+\tau) \rangle
   \label{eq:COORDisfa}
\end{equation}
which may be written in terms of the bead positions as
\begin{equation}
      g^{(1)}(\mathbf{q}, \tau) = \langle \sum_{i=1}^{N} \sum_{j=1}^{N}  \exp(-\imath \mathbf{q} \cdot (\mathbf{r}_{i}(t)-\mathbf{r}_{j}(t+\tau))).
    \label{eq:COORDisfb}
\end{equation}
By inspection, $g^{(1)}(\mathbf{q}, \tau)$ does not vanish by translational invariance. That's a statement about the statistical weight $\exp(- \beta (H(\mathbf{\Gamma})-A)$ in equation \ref{eq:COORDcorrelation}, namely $H(\mathbf{\Gamma})$ is only a function of the distances between pairs of atoms, and those distances are not changes by a displacement of the origin.

 The double sum is usefully divided into its self ($i=j$) and distinct ($i \neq j$) parts, these being denoted  $g^{(1s)}(\mathbf{q}, t)$ and $g^{(1d)}(\mathbf{q}, t)$, respectively.  Some experimental techniques measure $g^{(1)}(\mathbf{q}, t)$, while others measure $g^{(1s)}(\mathbf{q}, t)$.

It is sometimes asserted that the $a_{\mathbf{q}}(t)$ are uncorrelated, by which it is meant that
\begin{equation}
        \langle  a_{\mathbf{k}}(t)  a_{\mathbf{q}}(t) \rangle = \langle \sum_{i=1}^{N} \sum_{j=1}^{N}  \exp(-\imath \mathbf{k} \cdot \mathbf{r}_{i}(t)+ \imath \mathbf{q} \cdot   \mathbf{r}_{j}(t))) \rangle.
    \label{eq:COORDcrosscorr1}
\end{equation}
Indeed, the result $\langle  a_{\mathbf{k}}(t)  a_{\mathbf{q}}(t) \rangle = 0$ for $k + q \neq 0$ follows from translational invariance.

However, the three-mode cross-correlation function  $\langle  a_{\mathbf{k}}(t)  a_{\mathbf{q}}(t)  a_{\mathbf{p}}(t) \rangle$ may be non-zero if $\mathbf{k} +\mathbf{q} + \mathbf{p} = \mathbf{0}$.  When  $\mathbf{k} +\mathbf{q} + \mathbf{p} = \mathbf{0}$ is satisfied, the three-mode cross correlation function becomes $\langle  a_{\mathbf{k}}(t)  a_{\mathbf{q}}(t)  a_{-\mathbf{k}-\mathbf{q}}(t) \rangle$.  This function is proportional to the double spatial fourier transform of the three-point distribution function $g^{(3)}(\mathbf{r}_{1}, \mathbf{r}_{2}, \mathbf{r}_{3})$, namely
\begin{equation}
     \tilde{g}^{(3)}(\mathbf{k},\mathbf{q}) = \int_{V} d \mathbf{r}_{1}d \mathbf{r}_{2}d \mathbf{r}_{3} g^{(3)}(\mathbf{r}_{1}, \mathbf{r}_{2}, \mathbf{r}_{3}) \exp(\imath (\mathbf{r}_{2} - \mathbf{r}_{1}) \cdot \mathbf{k}) \exp(\imath (\mathbf{r}_{3} - \mathbf{r}_{1}) \cdot \mathbf{q}).
     \label{eq:COORDthreemode}
\end{equation}
$g^{(3)}(\mathbf{r}_{1}, \mathbf{r}_{2}, \mathbf{r}_{3})$ is the three-particle distribution function.  It gives the equilibrium average likelihood of finding three beads at relative positions $\mathbf{r}_{1}$, $\mathbf{r}_{2}$, and $\mathbf{r}_{3}$.

\section{Haar Wavelet Components \label{COORDwavelets}}

Wavelets are a relatively new mathematical form\cite{COORDhaar1910a,COORDdaubechies1992a}.  Like Fourier transforms, wavelet transforms can be described as a set of coordinates, amplitudes corresponding to those coordinates, and, potentially, wavelets as normal modes.  However, unlike Fourier or Rouse coordinates, wavelet coordinates provide a natural description for the dynamics of local regions of a polymer.  Wavelets are therefore potentially a natural form for describing polymer dynamics, though their applications have thusfar only had limited use in polymer or soft-matter physics (though note Phillies and Stott\cite{COORDphillies1995z,COORDphillies1995y} and Whitford and Phillies\cite{COORDwhitford2005a,COORDwhitford2005b}).  There are an extremely large number of sets of wavelet decompositions\cite{COORDdaubechies1992a}.  In the case at hand, a polymer is represented as a discrete, finite list of positions and momenta, so a discrete set of wavelet coordinates is most appropriate. As an example, the use of wavelets resembling Haar\cite{COORDhaar1910a} wavelets is shown.

The position coordinates of an $N$-bead polymer may be described as a set of $N$ bead positions.  The decomposition rearranges the coordinates into a series of average wavelet components $c$ and a set of difference wavelet components $d$.  The individual wavelet components are written $c(n,\alpha,j)(t)$ and $d(n,\alpha,j)(t)$.  Here $n$ is the wavelet decomposition level, $\alpha = (x,y,z)$ is the corresponding Cartesian coordinate, and $j$ labels the wavelet location along the polymer chain. Wavelets are generated iteratively, with each set of averaged positions being generated from the averaged components generated by the prior iteration. With each iteration, the number of $c(n,\alpha,j)(t)$ and of $d(n,\alpha,j)(t)$ generated by the iteration is half the number of $c(n-1,\alpha,j)(t)$ generated in the prior iteration.  The maximum value of $j$ depends on $N$.  For a $2^{m}$ bead polymer the upper limit on $j$, the number of components generated in a single iteration, is $2^{m-n}$ with $m-n \geq 0$.

The lowest-order Haar wavelet amplitudes are calculated from the bead positions as
\begin{eqnarray}
  c(1,\alpha,j) &=& (r_{\alpha, 2*j} + r_{\alpha, 2*j-1})/2  \\
  d(1,\alpha,j) &=& (r_{\alpha, 2*j} - r_{\alpha, 2*j-1})/2,
  \label{eq:COORDdef1}
\end{eqnarray}
The $c(1,\alpha,j)$ are the averaged position of pairs of adjacent beads, while the $d(1,\alpha,j)$ are the vectors from the average position to either of the beads in that average. The decomposition proceeds naturally if for some integer $m$ there are $2^{m}$ beads in the chain.

It would equally be possible to begin calculating the lowest-order wavelet amplitudes by using the $N-1$ interbead vectors, defined by $R_{\alpha,i} =  r_{\alpha, i+1} - r_{\alpha, i}$, in which case one could define
\begin{eqnarray}
  c(1,\alpha,j) &=& (R_{\alpha, 2*j} + R_{\alpha, 2*j-1})/2  \\
  d(1,\alpha,j) &=& (R_{\alpha, 2*j} - R_{\alpha, 2*j-1})/2.
  \label{eq:COORDdef2}
\end{eqnarray}
In this case, the decomposition proceeds naturally if there are $2^{m}$ vectors and therefore $2^{m}+1$ beads in the chain.

In either case, higher-order ($n > 1$) Haar components are generated iteratively via
\begin{eqnarray}
  c(n,\alpha,j) &=& (c(n-1,\alpha,2*j) + c(n-1, \alpha,2*j-1)/2 \\
  d(n,\alpha,j) &=& (c(n-1,\alpha,2*j) -  c(n-1, \alpha,2*j-1))/2
    \label{eq:COORDdef3}
\end{eqnarray}
At each iteration, the $c(n,\alpha,j)$ and  $d(n,\alpha,j)$ are calculated from the $ c(n-1,\alpha,j)$.

The $c(n,\alpha,j)$ refer to the average positions of longer and longer pieces of the chain.  Once calculated, the  $d(n,\alpha,j)$ play no further role in the iteration process, though they are required to reconstruct the original bead or vector positions from the $c(n,\alpha,j)(t)$ and the $d(n,\alpha,j)(t)$.

Time correlation functions of the wavelet components describe the dynamics of local chain segments having different lengths and positions along the chain.

\section{Models for Chain Dynamics \label{COORDmodels}}

We now turn to models of polymer dynamics.  The discussion is located here because one of the models leads naturally to Rouse coordinates and the Rouse transform.  We remind readers of  Likhtman's observations in his review \emph{Viscoelasticity and Molecular Rheology}\cite{COORDlikhtman2012a}, ``\emph{We note that often models are studied by theoreticians just because they are analytically solvable and used by experimentalists because of availability of analytic solutions.}'', and ``\emph{This coupling suggests that the Rouse mode description is not very useful for entangled polymers.}''

There are two important models for the significant motions of a single polymer chain in solution, these being the Kirkwood-Riseman model\cite{COORDkirkwood1948a} and the Rouse model\cite{COORDrouse1953a}.  The Kirkwood-Riseman model presents a natural description of chain motions. The Rouse model leads naturally to Rouse coordinates and Rouse normal modes.

Both models describe a polymer molecule as a line of beads, somewhat like the pearls in a necklace. In both models, all bead motions are massively overdamped, so that inertia is neglected.  The details of the bonds between adjoining beads differ in the two models. For each general model, one may separately construct forms that do or do not include hydrodynamic interactions between the beads of the polymer chain.

The Kirkwood-Risemann and Rouse models provide entirely contradictory descriptions as to how a polymer molecule contributes to the viscosity of a polymeric fluid. In the Rouse model, viscous dissipation arises from the internal modes of a polymer, in which the different parts of the molecule move with respect to each other and exert forces on each other.  Rouse-model polymers also have some whole-body motions, but these are taken not to contribute to viscous dissipation.  In the Kirkwood-Riseman model, viscous dissipation arises from polymeric whole-body motions, notably rotations, and forces between polymer beads and the surrounding solvent.   Kirkwood and Riseman recognized that polymers also have internal motions in which the relative positions of beads depend on time, but assumed that internal motions provide only secondary corrections, at least for the physical quantities in which they were interested.

The Kirkwood-Riseman model is less often considered than is the Rouse-Zimm model, in part because the Kirkwood-Riseman model is more demanding mathematically, in part because Kirkwood and Riseman use a less familiar notation, and in part because the presentation of the Kirkwood-Riseman model has not passed through the series of restatements seen for the Rouse model, restatements that have taken scholars from the elaborate presentation of Rouse to the simple and elegant presentations of Padding\cite{COORDpadding2005a}, Edwards and Doi\cite{COORDdoi1988a}, and Likhtman\cite{COORDlikhtman2012a}.

The next three sections discuss the Kirkwood-Riseman model, the Rouse model, and some peculiar features of the Rouse model and Rouse coordinates.

\subsection{Kirkwood-Risemann Model\label{ss:krm}}

We first consider the Kirkwood-Riseman\cite{COORDkirkwood1948a} model, restated here in more modern terms. Kirkwood and Riseman’s presentation was phrased entirely in Cartesian coordinates, which are also used here.  The Kirkwood-Riseman model describes a chain of $N$ beads connected by rigid links having length $b_{0}$. Adjoining links are separated by a rigid angle $\theta$. Successive three-bead planes define a torsion angle $\phi$. In the original model, the potential energy was taken to be independent of the angle $\phi$, but this assumption has only secondary consequences.  The effective bond length, the contribution of each link to the distance between distant beads, is therefore
\begin{equation}
   b = \left(\frac{1 + \langle \cos(\phi)\rangle}{1 - \langle \cos(\phi)\rangle}  \right)  \left(\frac{ 1 - \cos(\theta)}{1 + \cos(\theta)} \right) b_{0}.
   \label{eq:bb0}
\end{equation}

Kirkwood and Riseman assert that the probability distribution for the distance between two beads that are distant along the chain has a Gaussian form. For beads $\ell$ and $s$ that are well-separated, they supply several average values, notably
\begin{align}
    \langle \mid R_{\ell s}\mid^{2} \rangle  &= \mid \ell - s\mid b^{2} \\
    \langle \mid R_{0 \ell}\mid^{2} \rangle &= b^{2} \left(\frac{12 \ell^{2} + N^{2} - 2 N + 1}{12(N-1)}\right)\\
    \langle \mathbf{R}_{0\ell} \cdot  \mathbf{R}_{0s} \rangle &= \frac{b^{2}}{N-1}\left(\frac{\ell^{2} + s^{2}}{2} - \frac{N-1}{2}\mid\ell - s\mid + \frac{(N-1)^{2}}{2}  \right)\\
    \left\langle \frac{1}{R_{\ell s} }\right\rangle &= \frac{6}{\sqrt{\pi} b \mid \ell - s \mid^{1/2}}.
     \label{eq:distances}
\end{align}
Here beads $\ell$ and $s$ have locations $\mathbf{r}_{\ell}$ and $\mathbf{r}_{s}$, $\mathbf{R}_{\ell s} = \mathbf{r}_{s} -\mathbf{r}_{\ell}$ is the vector from bead $\ell$ to bead $s$, $R_{\ell s} = |\mathbf{R}_{\ell s}|$, and $\mathbf{r}_{0}$ is the location of the center of mass of the polymer, so that $\mathbf{R}_{0 \ell}$ is the vector from the polymer's center of mass to bead $\ell$.

The Kirkwood-Riseman model assumes that polymer beads have long range hydrodynamic interactions described by the Oseen tensor
\begin{equation}
     \mathbf{T}_{ij}(\mathbf{r}_{ij})  = \frac{1}{8 \pi \eta_{0} r_{ij}}(\mathbf{I} + \mathbf{\hat{r}}_{ij}\mathbf{\hat{r}}_{ij}),
    \label{eq:Oseen}
\end{equation}
which gives the fluid flow created at a point $\mathbf{r}_{j}$ by a force $\mathbf{F}_{i}$ applied to the solution at point $\mathbf{r}_{i}$.  The vector from point $i$ to point $j$ is $\mathbf{r}_{ij}$, with magnitude $r_{ij} = \mid \mathbf{r}_{ij}\mid$ and corresponding unit vector $\mathbf{\hat{r}}_{ij}= \mathbf{r}_{ij}/r_{ij}$.  Here $\eta_{0}$ is the solvent viscosity.  In eq.\ \ref{eq:Oseen} and its associated notation, there is no assumption that there is actually a polymer bead rather than solvent at the point $\mathbf{r}_{j}$.  Oseen's model treats the force as a point source, and assumes that the presence of the polymer has no effect on the solvent's innate viscosity, an assumption that is known experimentally to be incorrect\cite{COORDsolvent1}-\cite{COORDsolvent4}.

The fluid flow $\mathbf{v'}(\mathbf{r}_{j})$ induced by $\mathbf{F}_{i}$ at the point $\mathbf{r}_{j}$ is therefore
\begin{equation}
  \mathbf{v'}(\mathbf{r}_{j})  = \mathbf{T}_{ij}(\mathbf{r}_{ij}) \cdot \mathbf{F}_{i}({\bf r}_{i}).
    \label{eq:oseenflow}
\end{equation}

Within the model, the forces $\mathbf{F}_{i}$ arise because the beads move with respect to the fluid.  If a bead is stationary with respect to the local fluid flow, it exerts no force on the fluid.  The force exerted on the fluid by a bead $\ell$ is determined by the velocity $\mathbf{u}_{\ell}$ of the bead, the velocity $\mathbf{v}(\mathbf{r}_{\ell})$ that the fluid would have had, at the point $\mathbf{r}_{\ell}$, if the bead were not present, and the drag coefficient $f$ of a bead, namely
\begin{equation}
          \mathbf{F}_{i} = f(\mathbf{u}_{\ell} - \mathbf{v}(\mathbf{r}_{\ell}))
\label{eq:oseenforce}
\end{equation}
Because the beads are treated as points, a single bead exerts no torque on the surrounding fluid.

We now come to the dynamics of the polymer, as modelled by Kirkwood and Riseman.  The beads are taken to lie along a Gaussian chain, meaning that on the average the bead concentration declines with distance from the chain center of mass, the concentration depending on that distance as a gaussian in that distance. Kirkwood and Riseman describe polymer motions as a sum of whole-body translation and whole-body rotation. Within the model, the velocities of the individual beads are taken to be determined entirely by the time-dependent chain center-of-mass velocity $\mathbf{V}(t)$  and chain rotational velocity $\mathbf{\Omega}(t)$ as
\begin{equation}
           \mathbf{u}_{\ell}(t) = \mathbf{V}(t) + \mathbf{\Omega}(t) \times \mathbf{R}_{0\ell}
\label{eq:beadvelocity1}
\end{equation}
$\mathbf{u}_{\ell}$, as given by equation \ref{eq:oseenforce}, is the velocity that the bead would have had, if it were part of a rigid body that had translational velocity $\mathbf{V}$ and rotational velocity $\mathbf{\Omega}$.  The actual beads also have additional individual motions corresponding to the polymer internal modes, but these are neglected by Kirkwood and Riseman.

What forces act on a polymer chain? To calculate these, the model begins with the observation that, in the absence of external forces, the polymer's translational acceleration must over long times average nearly to zero. For the same reason, over long times the polymer's rotational acceleration must average nearly to zero. Polymer motions are heavily overdamped; the required 'long' time is actually quite short.

Newton's Third Law and its analog for rotational motion then tell us that if the average acceleration is zero, the force on the chain averaged over the same period must also be zero, and if the average angular acceleration is zero, the torque on the chain averaged over the same period must also be zero. These results are the zero-net-force and zero-net-torque conditions that are central to the Kirkwood-Riseman model.  The zero-net-force and zero-net-torque conditions determine the response of the polymer to an external force or to an external torque.

As an example of the Kirkwood-Riseman model and the effect of hydrodynamic interactions, we consider the drag coefficient (and hence the diffusion coefficient) of a polymer chain.  The analysis of Zwanzig\cite{COORDzwanzig1969a} is followed here.  Note that Kirkwood and Riseman took $\mathbf{F}_{j}$ to be the hydrodynamic force of the bead on the solvent, while Zwanzig takes $\mathbf{F}_{j}$ to be the force on the bead by the solvent, these forces being compelled by Newton’s Third Law to be equal in magnitude and opposite in direction, so the two papers have sign differences. We first consider a polymer chain whose beads have arbitrary velocities $\mathbf{u}_{\ell}$, while the fluid at $\mathbf{r}_{\ell}$ has an unperturbed velocity $\mathbf{v}_{\ell}^{0}$. The hydrodynamic interactions perturb the fluid flow at $\mathbf{r}_{\ell}$, so the actual fluid velocity at $\mathbf{r}_{\ell}$ is
\begin{equation}
      \mathbf{v}_{\ell} =  \mathbf{v}_{\ell}^{0} + \sum_{k \neq \ell =1}^{N} \mathbf{T}_{\ell k} \cdot \mathbf{F}_{k}.
      \label{eq:perturbedvelocity}
\end{equation}

However, the hydrodynamic force that a bead $k$ exerts on the solvent is
\begin{equation}
      \mathbf{F}_{k} = f (\mathbf{u}_{k} - \mathbf{v}_{k}),
      \label{eq:solventforce}
\end{equation}
$f$ being the drag coefficient of a single bead. Combining the above two equations,
\begin{equation}
      \mathbf{v}_{\ell} =  \mathbf{v}_{\ell}^{0} - f \sum_{k \neq \ell =1}^{N} \mathbf{T}_{\ell k} \cdot f (\mathbf{v}_{k} - \mathbf{u}_{k}) .
      \label{eq:perturbedvelocity2}
\end{equation}
Subtracting  $\mathbf{u}_{\ell}$ from each side of the equation,
\begin{equation}
      \mathbf{v}_{\ell} -\mathbf{u}_{\ell} =  \mathbf{v}_{\ell}^{0}-\mathbf{u}_{\ell} - f\sum_{k \neq \ell =1}^{N} \mathbf{T}_{\ell
k} \cdot f (\mathbf{v}_{k} - \mathbf{u}_{k}) .
      \label{eq:perturbedvelocity3}
\end{equation}
which allows us to write
\begin{equation}
      \mathbf{v}_{\ell}^{0} -\mathbf{u}_{\ell} =   f \sum_{k  =1}^{N} \mathbf{\mu}_{\ell k} \cdot (\mathbf{v}_{k} - \mathbf{u}_{k}) .
      \label{eq:perturbedvelocity4}
\end{equation}
This equation introduces a new matrix $\mathbf{\mu}$, the matrix being
\begin{equation}
           \mathbf{\mu}_{\ell k} = \frac{\mathbf{I} \delta_{\ell k}}{f} + \mathbf{T}_{\ell k},
     \label{eq:muzwanzigdef}
\end{equation}
where the rule $\mathbf{T}_{kk} = 0$ has been applied and where $\mathbf{I}$ is a $3 \times 3$ identity matrix.

Matrix inversion gives the $\mathbf{v}_{k} - \mathbf{u}_{k}$ in terms of the $\mathbf{v}_{\ell}^{0} -\mathbf{u}_{\ell}$ and the inverse $\mathbf{\mu}^{-1}$ of $\mathbf{\mu}$, namely
\begin{equation}
         \mathbf{v}_{k} - \mathbf{u}_{k} = f^{-1} \sum_{\ell=1}^{N} (\mathbf{\mu}^{-1})_{k\ell} \cdot (\mathbf{v}_{\ell}^{0} -\mathbf{u}_{\ell})
     \label{eq:actualvelocities}
\end{equation}
so the force on a bead $k$ due to its hydrodynamic interactions with the solvent becomes
\begin{equation}
      - \mathbf{F}_{k} \equiv f(\mathbf{v}_{k} - \mathbf{u}_{k}) = - \sum_{\ell=1}^{N} (\mathbf{\mu}^{-1})_{k\ell} \cdot (\mathbf{v}_{\ell}^{0} -\mathbf{u}_{\ell}).
      \label{eq:Fkvalue}
\end{equation}
The minus sign appears because $\mathbf{F}_{k}$ is the force of the bead on the solvent, not the force of the solvent on the bead.

The drag coefficient $f_{c}$ of the polymer chain is obtained by choosing all bead velocities to be equal to $\mathbf{u}_{0}$, the unperturbed fluid velocity $\mathbf{v}_{0}$ to be zero, and calculating the total of the drag forces on all beads of the chain, leading to
\begin{equation}
      - \sum_{k=1}^{N} \mathbf{F}_{k} \equiv  f_{c} \mathbf{u}_{0} = \sum_{k=1}^{N}  \sum{\ell=1}^{N} (\mathbf{\mu}^{-1})_{k\ell} \cdot \mathbf{u}_{\ell}
      \label{eq:dragcoefficient}
\end{equation}
with $\mathbf{u}_{\ell} = \mathbf{u}_{0} \ \forall \ \ell$.

In averaging equation \ref{eq:dragcoefficient} over all bead configurations, to determine the beads ensemble-average drag coefficient $\langle f \rangle$, Kirkwood and Riseman advanced by replacing all bead-bead distances with their average values, i.e., they implicitly wrote
\begin{equation}
      \langle \mathbf{\mu}^{-1})_{k\ell}(\Gamma) \rangle =  \mathbf{\mu}^{-1})_{k\ell}(\langle \Gamma  \rangle)
\end{equation}
where $\Gamma$ is the list of all coordinates on which $\mathbf{\mu}^{-1})_{k\ell}$ depends.

Bead-bead hydrodynamic interactions as described by the Oseen tensor thus perturb the drag coefficient of the whole chain. Further calculations of the same sort allow calculation of the contribution of a Kirkwood-Riseman model chain to the solution viscosity.

\section{Rouse Components \label{COORDrouse}}

We now turn to Rouse coordinates, Rouse transforms, and Rouse modes. The Rouse transformation may be recognized as a discrete Fourier transform.  In a Rouse transform, the transform variables are the label $i$ identifying the bead and a label $p$ identifying the Rouse coordinate.  $i$ is the formal analog of the position variable $x$.  The transform variable $p$ is analogous to the wavevector $k$.  In the Rouse transform, the complex exponential of the continuous Fourier transform is replaced by the exponential's real part.  The variable $i$ is not necessarily abstract.  In the Rouse model, $i$ identifies which beads are linked to which other beads, a bead $i$ being linked to beads $i \pm 1$.  The Cartesian position coordinates $x_{i}$, $y_{i}$, and $z_{i}$ of each bead are to be interpreted as the values of a function $\mathbf{R}_{i}(t)$ of the variable $i$.

For an $N+1$ bead polymer chain, there are $3N+3$ position coordinates and correspondingly $3N+3$ Rouse coordinates.  We denote the amplitudes of the Rouse coordinates by $X_{\alpha p}$.  Here $\alpha$ labels the Cartesian coordinate that corresponds to the transform. It is of course possible to replace $\{X_{xp}(t), X_{yp}(t), X_{zp}(t)\}$ by a single three-vector $\mathbf{X}(t)$, but that would obscure the issue that the Rouse model's solutions, which are three-dimensional, naturally partition into three one-dimensional parts.

The three $X_{\alpha 0}$ give the position of the polymer's center of mass.   The $N$ Rouse amplitudes $X_{xp}(t)$ for the $x$-direction coordinates, with $p \in (1, N)$, are computed from the bead $x$-coordinates $x_{i}(t)$ via the Rouse transform
\begin{equation}
    \label{eq:COORDmodeamplitudes}
     X_{xp}(t)  = \frac{1}{N+1} \sum_{i=0}^{N} x_{i}(t) \cos\left(\frac{p \pi (i+1/2)}{N+1}\right).
\end{equation}
Because the polymer bead positions $x_{i}(t)$ change with time, the $X_{xp}(t)$ are time-dependent. Entirely similar equations give the Rouse amplitudes $X_{yp}$ and $X_{zp}$ for the $y$- and $z$-components of the particle positions.

The vector form of this equation is
\begin{equation}
    \label{eq:COORDmodeamplitudesvec}
     \mathbf{X}_{p}(t)  = \frac{1}{N+1} \sum_{i=0}^{N} \mathbf{R}_{i}(t) \cos\left(\frac{p \pi (i+1/2)}{N+1}\right).
\end{equation}
The corresponding set of inverse equations give the $x$-coordinates $x_{i}(t)$ of the polymer beads in terms of the Rouse amplitudes, namely
\begin{equation}
   \label{eq:COORDbeadpositions}
   x_{i}(t) = X_{x0}(t) + 2 \sum_{p=1}^{N} X_{xp}(t) \cos\left(\frac{p \pi (i+1/2)}{N+1}   \right).
\end{equation}
Totally similar equations give the $y_{i}(t)$ and $z_{i}(t)$ in terms of the $X_{yn}(t)$ and the $X_{zn}(t)$, respectively.

The Cartesian-space representation of each Rouse coordinate follows from eqn.\ \ref{eq:COORDbeadpositions} on setting all but one of the $X_{xp}(t)$ to zero, and setting the final $X_{xp}(t) = 1$, giving for the Cartesian representation of $X_{xp}(t)$ an $N+1$ list of (generally) non-zero positions\cite{COORDpadding2005a}
\begin{equation}
     {x_{i}(t)}  = {\cos\left(\frac{p \pi (i+1/2)}{N+1}   \right)}.
     \label{eq:COORDrouseincartesian}
\end{equation}
In Cartesian representation, the $y$- and $z$-components of $X_{xp}(t)$ are all equal to zero, and similarly for $X_{yp}(t)$ and $X_{zp}(t)$. .

The positions of the beads in a polymer may be represented either as a list of $3N+3$ values for their $3N+3$ Cartesian coordinates or as a list of values for $3N+3$ amplitudes for their Rouse coordinates.   Either list is complete; the two lists each contain exactly the same information about the positions of the beads.  If one list is known, the other follows immediately. Equations \ref{eq:COORDmodeamplitudes}-\ref{eq:COORDbeadpositions} are the linear transformations that link the two sets of coordinates.  These equations are a purely mathematical result and have no intrinsic physical content.

\section{The Rouse and Zimm Models\label{eq:rzm}}

This Section presents the Rouse and Zimm models for the motions of an isolated polymer molecule in a simple Newtonian solvent. I follow here the recent treatments of Padding\cite{COORDpadding2005a}, Doi and Edwards\cite{COORDdoi1988a}, and Likhtman\cite{COORDlikhtman2012a} .

Rouse\cite{COORDrouse1953a} proposed a simple model for an isolated polymer chain, in the form of a series of beads linked by springs. The polymer is approximated as a line of $N+1$ beads whose positions are $(\mathbf{R}_{0}, \mathbf{R}_{1}, \mathbf{R}_{2}, \ldots \mathbf{R}_{N})$, respectively.  The beads do not have excluded-volume interactions, but they do exchange frictional forces with the solvent.  The frictional force exerted on a bead $i$ by the solvent is $- f \mathbf{v}_{i}$, where $\mathbf{v}_{i}$ is bead $i$'s velocity with respect to the solvent. In Rouse's original model, there were no hydrodynamic interactions between beads. Bead motions are heavily overdamped, so bead inertia is neglected within the model.

Each spring is connected at each end to a bead.  The springs are Hookean, and have zero length when no outside forces are applied to them.   The springs are hydrodynamically inert, so that they exert no forces on the solvent. An amended Rousian model due to Zimm\cite{COORDzimm1956a} follows the Kirkwood-Riseman model\cite{COORDkirkwood1948a} in inserting hydrodynamic interactions between beads at the level of the Oseen Tensor.

It is noteworthy that Rouse appears to have been the first to propose that a polymer chain in dilute solution responds to an applied shear via an affine deformation\cite{COORDrouse1953a}, as opposed to responding to an applied shear via rotation, as predicted by the Kirkwood-Riseman model\cite{COORDkirkwood1948a}. Rouse's proposal for affine deformation, which for single chains is rejected by simulations\cite{COORDphillies2018a}, is a central feature of the deGennes\cite{COORDdegennes1979a} description of polymeric fluid dynamics.

In the Rouse model, the distribution $P(r)$ of distances $r$  between a pair of neighboring beads is taken to be a Gaussian $P(r) \sim \exp(- a r^{2})$.  Corresponding to this distribution, there is a matching potential of average force $W(r) = - k_{B}T \ln(P(r))$, with $k_{B}$ being Boltzmann's constant and $T$ being the absolute temperature. $W(r)$ is therefore a quadratic in $r$, namely
\begin{equation}
      W(r) = \frac{1}{2} k r^{2},
      \label{eq:rousepoaf}
\end{equation}
$k$ being a constant. Within the model, the 'spring constant' $k$ and the mean-square bead separation $b^{2}$ are related by
\begin{equation}
      \frac{1}{2}kb^{2} = \frac{3}{2} k_{B}T,
      \label{eq:rousekdefinition}
\end{equation}
which may be recognized as the equipartition theorem.  Corresponding to the potential of average force, each linked pair of beads $(i, i+1)$ is subject to an attractive Hooke's-Law force having magnitude $|k (\mathbf{R}_{i+1}-\mathbf{R}_{i})|$.  The force vanishes when the beads are at the same location.

In the Rouse model, the equations of motion of the beads are therefore
\[
f \frac{d \mathbf{R}_{i}}{dt} =
\begin{cases}
      -\frac{3 k_{B} T}{b^{2}} (\mathbf{R}_{1} - \mathbf{R}_{0}) + \mathbf{f}_{0}(t), &\text{if $i=0$;}\\
 -\frac{3 k_{B} T}{b^{2}} (2 \mathbf{R}_{i} - \mathbf{R}_{i-1} - \mathbf{R}_{i+1})  + \mathbf{f}_{i}(t), &\text{if $1 \leq i \leq N-1$;}\\
 -\frac{3 k_{B} T}{b^{2}} (\mathbf{R}_{N} - \mathbf{R}_{N-1}) + \mathbf{f}_{N}(t), &\text{if $i=N$.}
\end{cases}
\]
$\mathbf{f}_{i}(t)$ is the thermal force on bead $i$, physically arising from interactions of the bead with the solvent. $\mathbf{f}_{i}(t)$ and the friction factor $f$ are not independent; they are linked by a fluctuation-dissipation theorem. In the Rouse model, beads do not exert hydrodynamic forces on each other; correspondingly, the random forces on different beads are not correlated with each other. The above three equations are an elaborate way to write that inertia has been suppressed, so the total force on each bead must vanish: the mechanical and hydrodynamic forces on each bead must sum to zero.

The above equations are a set of $N+1$ vector equations and therefore $3N+3$ scalar equations. Within this model, equations corresponding to different Cartesian axes are uncoupled. Noting that the $x$-component direction cosine for the vector between beads $i$ and $i+1$ is  $(x_{i+1}-x_{i})/(\mid (\mathbf{R}_{i+1}-\mathbf{R}_{i})\mid )$, and similarly for the $y$-component and the $z$-component, the vector equations for the interior beads may be replaced by three sets of scalar equations, viz.,
\begin{equation}
     f \frac{d x_{i}(t)}{dt} = -k(2 x_{i} - x_{i-1}  - x_{i+1}) + F_{ix} (t)
     \label{eq:rousemostbeadsx}
\end{equation}
for the $x$ coordinates, and matching sets of equations for the $y$ and $z$ coordinates.  Here $x_{i}$ and $F_{ix}$ are the $x$ coordinate of particle $i$ and the $x$ component of the thermal force on particle $i$.

For bead 0, the corresponding equation is
\begin{equation}
     f \frac{d x_{0}(t)}{dt} = k(x_{1} - x_{0}) + F_{0x} (t)
     \label{eq:rouseendbeadsx}
\end{equation}
and correspondingly for bead $N$. The Rouse model thus yields a set of $3(N+1)$ coupled first-order linear differential equations.

However, as noted by Rouse, the equations for the $x$, $y$, and $z$ coordinates are entirely uncoupled, and are the same except for coordinate label, so one only needs to solve a set of $N+1$ coupled linear differential equations, and that once, to know the complete solution.  It should be stressed, however, that the Rouse model \emph{is not a one-dimensional model. It is a three-dimensional model.}  The Rouse model describes bead motions in all three Cartesian directions.  We largely discuss the solutions for the $x$ coordinate, but the solutions for the $y$ and $z$ coordinate are the same except for the coordinate label.

The $x$-coordinate equations are sometimes written in a generalized vector form, the vector being $\mathbf{X}(t) = (x_{0}(t), x_{1}(t),\ldots, x_{N}(t))$, namely
\begin{equation}
        \mathbf{H} \cdot \frac{d \mathbf{X}(t)}{dt} = - \mathbf{A} \cdot \mathbf{X}(t).
      \label{eq:rousexvectorform}
\end{equation}
In the Rouse model the hydrodynamic interaction matrix $\mathbf{H}$ is $\mathbf{H} = f \mathbf{I}$, $\mathbf{I}$ being the identity matrix. In the Zimm model $\mathbf{H}$ has a more complex form reflecting bead-bead hydrodynamic interactions.  The interaction matrix $\mathbf{A}$ for a linear chain is
\begin{equation}
      \mathbf{A} = \begin{Bmatrix} 1 & -1 & 0 &0 & \ldots & 0 & 0 & 0 \\  -1 & 2 & -1 &0 & \ldots & 0 & 0 & 0 \\ 0 &  - 1 & 2 & - 1  & \ldots & 0 & 0 & 0  \\  & & & & & & &  \\  0 & 0 & 0 &0 & \ldots & -1 & 2 & -1  \\  0 & 0 & 0 &0 & \ldots & 0 & -1 & 1 \end{Bmatrix}
      \label{eq:rouseinteractionmatrix}
\end{equation}
Equation \ref{eq:rousexvectorform} is a set of $N+1$ coupled linear first-order differential equations. These equations were solved by Rouse. Its solutions are a set of $N+1$ eigenmodes $X_{p}$ whose forms are given by the Rouse coordinates discussed above.

At any time $t$, the normal mode amplitudes $X_{p}(t)$ for the $x$-coordinate modes are determined by a projection of the $x$-coordinates $x_{i}$ onto the Rouse transform, namely
\begin{equation}
      X_{p}(t) =  \frac{1}{N+1} \sum_{n=0}^{N} x_{i}(t) \cos\left(\frac{i \pi (i+1/2)}{N+1}\right).
      \label{eq:rouseamplitudes}
\end{equation}
Corresponding equations give the amplitudes of the $y$ and $z$ normal mode coordinates in terms of the $y_{i}$ and $z_{i}$, respectively.

The displacements $x_{i}$ of the individual atoms can be computed from the amplitudes of the normal modes as
\begin{equation}
     x_{i}(t) = X_{0}(t) + 2 \sum^{N}_{p=1} X_{p}(t) \cos\left(\frac{p \pi (i+1/2)}{N}   \right).
     \label{COORDdisplacements}
\end{equation}
Equations identical to eq.\ \ref{COORDdisplacements}, except for the coordinate label and the values of the normal mode amplitudes, describe the $y_{i}$ and the $z_{i}$.

The equations of motion for the $\mathbf{X}_{p}$ follow from the Rouse equations of motion as
\begin{equation}\label{eq:ROUSEexpeqnmotion}
   f_{p} \frac{d \mathbf{X}_{p}}{dt}  = - k_{p}  \mathbf{X}_{p} +\mathbf{f}_{p}(t).
\end{equation}
The above is a linear differential equation with constant coefficients and a source term; the solution is a decaying exponential.

For $p >0$ one has $f_{p} = 2 (N+1) f$ and
\begin{equation}
   k_{p} = \frac{24 k_{B} T (N+1)}{b^{2}} \sin^{2}\left(\frac{p \pi}{2(N+1)} \right).
   \label{eq:ROUSEkpdef}
\end{equation}
and for $p=0$ one has $f_{p} = (N+1) f$ and $k_{0} = 0$.  Here $\mathbf{f}_{p}(t)$ is the $p^{\rm th}$ component of the random force, which in the Rouse satisfies
\begin{equation}\label{eq:ROUSEpthrandomforce}
    \langle f_{p\alpha}(t) f_{q\beta}(\tau) \rangle = 2 k_{B} T f_{p} \delta_{pq} \delta_{\alpha \beta} \delta(t-\tau)
\end{equation}
where $\alpha$ and $\beta$ denote Cartesian components and $\delta$ is the Kronecker or Dirac delta function.

Finally, for $p > 0$ the time correlation function for the Rouse amplitudes is
\begin{equation}\label{eq:ROUSExptcf}
   \langle X_{p\alpha}(t) X_{q\beta}(0) \rangle = \langle (X_{p\alpha}(0))^{2} \rangle \delta_{pq} \delta_{\alpha\beta}  \exp(-t/\tau_{p}),
\end{equation}
where
\begin{equation}\label{eq:ROUSEtaupotherdef}
   \tau_{p}  = \frac{f b^{2}}{12 k_{B} T} \sin^{-2}\left(\frac{p \pi}{2(N+1)}\right) \approx \frac{f b^{2} (N+1)^{2}}{3 p^{2} \pi^{2} k_{B} T}
\end{equation}
and
\begin{equation}\label{eq:ROUSEmeansquareX}
   \langle (X_{p\alpha}(0))^{2} \rangle =  \frac{b^{2}}{8(N+1}  \sin^{-2}\left(\frac{p \pi}{2(N+1)}\right) \approx \frac{(N=1)b^{2}}{2 \pi^{2} p^{2}},
\end{equation}
the approximation being valid to within 1\% for $N > 10$.  The $X_{p\alpha}$ are predicted to have independent Gaussian random distributions, so all higher moments of $X_{p\alpha}$ can be calculated from $\langle (X_{p\alpha}(0))^{2} \rangle$.

As the closing and central point, the Rouse model for an $N+1$ bead polymer chain has as solutions a set of normal modes, namely three modes with eigenvalue zero and $3N$ modes with non-zero eigenvalues in which the beads move with respect to each other.  The modes: (i) are orthogonal (cf.\ eqn.\ \ref{eq:ROUSExptcf}, (ii) have as time correlation functions decaying as simple exponentials (cf.\ eqn.\ \ref{eq:ROUSExptcf}, (iii) have relaxation times that scale (for larger $(N+1)/p$) as $(N+1)^{2}/p^{2}$ (cf.\ eqn.\ \ref{eq:ROUSEtaupotherdef}), and (iv) have amplitudes whose mean-square values follow eqn.\ \ref{eq:ROUSEmeansquareX}.  Furthermore, if one calculates the bead velocities $d \mathbf{R}_{i}(t)/dt$ for each mode, one finds that the velocities are always directed exactly opposite to $\mathbf{R}_{i}(t)$, i.e., they are always directed at the chain center-of-mass.

Comparing these different sets of coordinates, the $a_{\mathbf{q}}(t)$ and the $X_{p}(t)$ are both global variables, each depending in general on the relative positions of all the beads in a chain.  In contrast, each Haar wavelet $c(n,\alpha,j)$ or $d(n,\alpha,j)$ is a localized variable; in general each refers to the behavior of a limited part of a polymer chain.

\end{document}